\documentclass{emulateapj}
\usepackage{natbib,graphicx,url}
\slugcomment{\sc Received 2010 June 21; accepted 2010 July 16; published 2010 July 30}
\shorttitle{1ES 1959+650: a multifrequency campaign}
\shortauthors{Bottacini et al.}

\begin{document}


\title{Probing the transition between the synchrotron and inverse-Compton spectral components 
of 1ES 1959+650}

\author{E. Bottacini\altaffilmark{1}, M.B\"ottcher\altaffilmark{2}, P. Schady\altaffilmark{1},
A. Rau\altaffilmark{1}, X.-L. Zhang\altaffilmark{1}, M. Ajello\altaffilmark{3}, C. Fendt\altaffilmark{4},
J. Greiner\altaffilmark{1}}

\altaffiltext{1}{Max-Planck Institut f\"ur Extraterrestrische Physik, Giessenbachstrasse,  85741 
		 Garching, Germany.}
\altaffiltext{2}{Astrophysical Institute, Department of Physics and Astronomy, Ohio University, 
     Athens, OH 45701, USA.}
\altaffiltext{3}{Stanford Linear Accelerator Center/KIPAC, 2572 Sand Hill Road, Menlo Park, 
     CA 91125, USA.}
\altaffiltext{4}{Max-Planck Institut f\"ur Astronomie, K\"onigstuhl 17,  69117 
		 Heidelberg, Germany}

\begin{abstract}
1ES~1959+650 is one of the most remarkable high-peaked BL Lacertae objects (HBL). 
In 2002 it exhibited a TeV $\gamma$--ray flare without a similar brightening of the 
synchrotron component at lower energies. This orphan TeV flare remained a mystery.
We present the results of a multifrequency campaign, triggered by the {\it INTEGRAL} IBIS detection 
of 1ES~1959+650. Our data range from the optical to hard X-ray energies, thus covering the 
synchrotron and inverse-Compton components simultaneously.
We observed the source with {\it INTEGRAL}, {\it Swift}/XRT, and UV-Optical Telescope, and nearly simultaneously 
with ground-based optical telescope.
The steep spectral component at X-ray 
energies is most likely due to synchrotron emission, while at soft $\gamma$-ray energies the 
hard spectral index may be interpreted as the onset of the high-energy component of 
the blazar spectral energy distribution (SED). This is the first clear measurement 
of a concave X-ray -- soft $\gamma$-ray spectrum for an HBL. 
The SED can be well modeled with a leptonic synchrotron- self-Compton model. When the SED is 
fitted this model requires a very hard electron spectral 
index of $q$ $\sim$ 1.85, possibly indicating the relevance of second-order Fermi acceleration. 
\end{abstract}

{\it Online-only material:} color figures

\section{Introduction}

Blazars are very bright from radio 
frequencies to gamma-ray energies \citep{ulrich97} due to the alignment of their jets 
with respect to the line of sight of the observer \citep{antonucci93,urry95}. Relativistic 
beaming in these outflows \citep{rees66} enhances the observed flux and shortens the variability 
timescales. The spectral energy distribution (SED) typically shows a non-thermal two-component 
structure. The low-energy emission is generally believed to be produced through synchrotron 
radiation of relativistic electrons. In contrast, the nature of the high-energy component 
is still being debated. A possible explanation is that the same electron population that 
is responsible for the low-energy component generates the high-energy component through 
Compton scattering \citep{sikora94, dermer93}. Alternative 
explanations are proton-initiated cascades \citep{mannheim93} or proton-synchrotron 
emission from ultrarelativistic protons \citep{muecke03}.

The position of the synchrotron peak defines two classes of BL Lac objects: high 
frequency peaked BL Lac objects (HBL, peak at UV--X-ray frequencies) and Low frequency 
peaked BL Lac objects (LBL, peak in the IR--optical band). 1ES~1959+650, at z = 0.046 
\citep{veron-cetty06}, is one of the best-studied member of the former group. It 
was first detected at X-rays during the Slew Survey of {\it Einstein}--IPC \citep{elvis92}. 
Further observations detected the object at X-rays with {\it ROSAT} and {\it BeppoSAX} 
\citep{beckmann02}. 
Based on the X-ray/radio versus X-ray/optical color--color diagram the source was classified as BL Lac 
object by \citet{schachter93}. The blazar was also detected at $\gamma$-rays by EGRET 
\citep{hartman99}. Very recently, the source was included in the {\it Fermi} first catalog 
of active galactic nucleus (AGN;\citealt{abdo10}). 1ES~1959+650 was observed several times at TeV energies 
\citep{aharonian03}. On 2002 June 4, followed by the detection of a strong TeV $\gamma$-ray
flare of the source with the 10m Whipple \v{C}erenkov Telescope, Target of Opportunity observations 
in optical and X-rays were performed. Despite an increased activity in the $\gamma$--rays, no flux 
variation was detected in X-rays \citep{krawczynski04} during simultaneous observations by 
the {\it Rossi X-Ray Timing Explorer (RXTE)}. As a
possible explanation, a hadronic synchrotron-mirror model has been suggested by 
\citet{boettcher05}, but this suffers from rather extreme energy requirements 
\citep{boettcher07}. Hence, the origin of the orphan TeV flare is at present not 
understood. It is therefore important to monitor the synchrotron and inverse-Compton 
components simultaneously.

In this Letter, we report on a multiwavelength campaign on 1ES~1959+650 ranging from 
the optical to hard X-ray energies.  In Section 2, we describe the observations of the 
multiwavelength campaign, and in Section 3 its results. In Section 4, we summarize our main
conclusions. 
\section{Observations}

\begin{table*}
\begin{minipage}[]{\textwidth}
\begin{center}
\caption{Log of {\it INTEGRAL}/IBIS, {\it Swift}/UVOT and Palomar 60- inch Observations. \label{tab:uvot}}
\begin{tabular}{llccccc}
\hline\hline
Instrument & Start Date UTC & Exp (s) & 	Band  & mag & Err & Flux$^{b}$\\
\hline
UVOT & 2007-Nov-16 02:34:55 & 1069 & $u$ & 14.96 & 0.02 & -\\
UVOT & 2007-Nov-23 03:00:46 & 479 & $uvw1$ & 14.98 & 0.03 & -\\
UVOT & 2007-Nov-23 04:34:45 & 539 & $uvw1$ & 15.09 & 0.02 & -\\
UVOT & 2007-Nov-30 17:54:42 & 98 & $uvw2$ & 15.33 & 0.03 & -\\
UVOT & 2007-Nov-30 17:56:10 & 165 & $uvw1$ & 15.19 & 0.03 & -\\
UVOT & 2007-Nov-30 17:58:55 & 388 & $u$ & 14.78 & 0.02 & -\\
UVOT & 2007-Nov-30 18:05:23 & 318 & $b$ & 14.63 & 0.02 & -\\
60 & 2007-Dec-03 01:56:02 & 30 &  $g$  &     15.28 &   0.03 & - \\
60 & 2007-Dec-04 01:56:33 & 30 &  $g$  &     15.29 &   0.03 & - \\
60 & 2007-Dec-05 01:54:03 & 30 &  $g$  &     15.32 &   0.03 & - \\
60 & 2007-Dec-11 02:25:35 & 30 &  $g$  &     15.38 &   0.03 & - \\
ISGRI & 2007-Nov-24 16:51:47 (624) & 55000  & - & - & - & 4.5$^{2.5}_{2.6}$\\
ISGRI & 2007-Nov-25 19:47:58 (625) & 198000 & - & - & - & 3.8$^{1.3}_{1.4}$\\
ISGRI & 2007-Nov-28 20:16:36 (626) & 198000 & - & - & - & 5.3$^{1.3}_{1.3}$\\
ISGRI & 2007-Dec-01 19:19:58 (627) & 208000 & - & - & - & 2.6$^{1.5}_{1.4}$\\ 
\hline
\end{tabular}
\end{center}
\end{minipage}
$^{a}$For ISGRI observation, revolution number is reported in
$^{b}$fluxes that are computed in the 20 -- 40 keV energy range in units of 10$^{-11}$ erg cm$^{-2}$ s$^{-1}$.
\end{table*}

\subsection{The 2007 multiwavelength campaign}
1ES 1959+650 was monitored by {\it INTEGRAL} (from 2007 November 27$^{th}$ to 2007 December 1$^{st}$ 2007 ) 
during the North Ecliptic Pole Key Program 
(proposal id: 0531000). The source was detected by IBIS/ISGRI \citep{bottacini07} in 
the 20 -- 40 keV band in an active state at a significance of 7.2 $\sigma$. Simultaneous X-ray 
and UV observations were obtained as a Target of Opportunity with {\it Swift} on 2007 November
30$^{th}$. From December 3-11, the Palomar 
(60) telescope performed optical photometry. Details of the multiwavelength campaign 
can be found in Table~\ref{tab:uvot}.
\subsubsection{UV to Optical Observations}
{\it UVOT.} 1ES~1950+650 was observed in the $u$, $b$, $uvw1$, and $uvw2$ bands (having central
wavelengths in units of {\mbox{{\AA}}} at 3465, 4392, 2600, 1928 respectively) with the UV-Optical 
Telescope (UVOT; \citealt{roming05}) on board the {\it Swift} satellite (Table~\ref{tab:uvot}). 
We used the standard pipeline reduced image products, co-added and exposure corrected within the 
XIMAGE\footnote{See http://heasarc.gsfc.nasa.gov/docs/xanadu/ximage/ximage.html} environment. 
For the photometry we used the standard 5$^{\prime \prime}$ aperture.
The reddening correction of $E(B-V) = 0.177$ was applied according to \citet{schlegel98} and 
\citet{cardelli89}.

{\it Palomar.} $g$-band (central wavelength 5240 {\mbox{{\AA}}})
 observations were carried out with the robotic 
Palomar 60- inch telescope \citep{cenko06} (see Table ~\ref{tab:uvot}). 
Data were reduced with standard IRAF routines. 
Photometric calibration was performed relative to the USNO-B1
catalog,\footnote{See http://www.nofs.navy.mil/data/fchpix} which leads to a systematic 
contribution to the photometric uncertainties of $\sim$ 0.5 mag.

\subsection{X-ray observations}
{\it INTEGRAL}. The {\it INTEGRAL} satellite observed the source during the North Ecliptic Pole Key 
Program with its imager IBIS/ISGRI \citep{lebrun03} which operates in the 17 -- 1000~keV 
range (Table~\ref{tab:uvot}). The observations were performed with a rectangular 5 $\times$ 5 
dithering pattern. We used the standard Off-line Science Analysis 
(OSA; \citealt{courvoisier03}) software version 7.0 for the ISGRI analysis. Most recent 
matrices available for standard software (isgri\_arf\_rsp\_0025.fits) were used for 
spectral analysis. We used 83 science 
windows for a total amount of 661 ks of exposure time. Data screening was 
performed according to the median count rate with respect to each science window and its 
distribution. After data cleaning the effective exposure on the source is 466 ks. 
{\it INTEGRAL}/SPI upper limits were obtained with the SPIMODFIT
software, which is available in OSA. It performs spectral model fitting for point sources and diffuse 
emission based on the maximum likelihood method as described in \citet{petry09}.
Due to the adopted dithering pattern, 1ES~1959+650 is outside the field of view of {\it INTEGRAL}'s 
Joint European Monitor for X rays (JEM-X; \citealt{lund03}) for at least 77~\% of the observation time.
For the same reason the Optical Monitoring Camera (OMC; \citealt{mas-hesse03}) could not point to 
the source. Therefore, no flux measurement could be obtained from these instruments.

{\it Swift/XRT.} The X-Ray Telescope (XRT; \citealt{burrows05}) on board {\it Swift} observed 
the blazar in 2007 November (Start date: UTC 2007 November 30 11:57:33) for an exposure time
of 1283 s. Data processing, screening and filtering were done using the FTOOL 
xrtpipeline included in the HEAsoft 6.3 distribution.
\section{Results}

\begin{table*}
\begin{minipage}[]{\textwidth}
\begin{center}
\caption{X-ray spectral fit results.\label{tab:fit}}
\begin{tabular}{ccccccccc}
\hline\hline
Inst. & Fit Mo & N$_{H}$ & $a$ & $b$ & E$_{b}$ & Norm & $\chi$ $^{2}$/dof & Flux$^{a}$ \\
(1) & (2) & (3) & (4) & (5) & (6) & (7) & (8) & (9)\\
\hline
XRT & ab pl & 0.17$^{0.04}_{0.03}$ & 2.1$^{0.1}_{0.1}$ & - & - & 4.1$^{0.4}_{0.3}$ & 178/192 & 1.09$^{0.02}_{0.02}$\\
XRT & ab pl & fixed                & 2.0$^{0.1}_{0.1}$ & - & - & 3.3$^{0.1}_{0.1}$ & 196/193 & 1.09$^{0.02}_{0.02}$\\
ISGRI & pl & - & 1.2$^{0.8}_{0.7}$ & & - & 7.8$^{2.6}_{7.6}$ & 0.6/3 & 1.4$^{0.3}_{1.2}$ \\
joint & ab bkn & fixed & 1.9$^{0.1}_{0.1}$ & 7.6$^{1.0}_{0.5}$ & 7.2$^{3.8}_{0.1}$ & 3.3$^{0.1}_{0.1}$ & 200/198 & - \\
joint & log-par & fixed & 1.3$^{0.1}_{0.2}$ & 0.6$^{0.1}_{0.1}$ & - & 3.4$^{0.1}_{0.2}$ & 323/199 & - \\
\hline
\end{tabular}
\end{center}
\end{minipage}
Explanation of columns: 
(1)instruments; 
(2)fit model; 
(3)column density (units of 10$^{22}$ cm$^{-2}$); 
(4)spectral index; 
(5)spectral index after break;  
(6)break energy; 
(7)normalization (units of 10$^{-4}$ ph keV$^{-1}$ cm$^{-2}$);
(8)statistical fit result;
(9)flux units are 10$^{-8}$ erg cm$^{-2}$ s$^{-1}$ for XRT and joint spectra and 
10$^{-11}$ erg cm$^{-2}$ s$^{-1}$ for IBIS/ISGRI spectrum;
$^{a}$fluxes are computed in the 0.6 -- 6.0 keV and 20 -- 60 keV energy ranges for the XRT and ISGRI respectively.
\end{table*}

\begin{figure}
\plotone{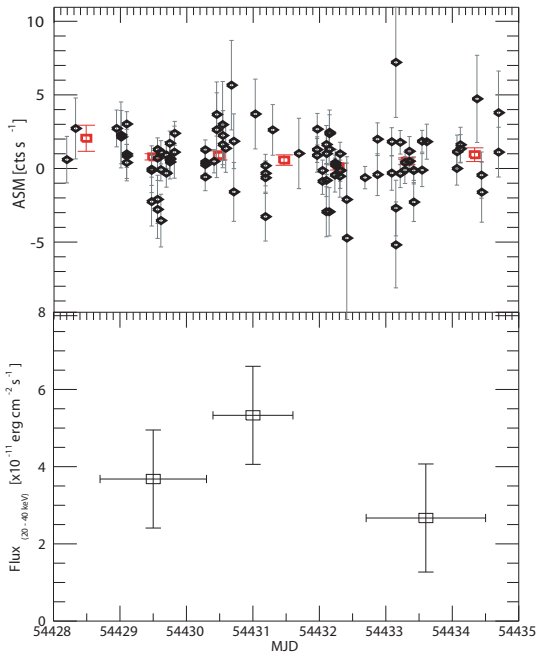}
\caption{Upper panel: ASM light curve of the source in the 1.5 -- 10~keV energy
band. Black diamond data points and red rectangles represent the dwell-by-dwell 90\,s exposure
and the one-day average light curves respectively.
Any flaring activity can be excluded. Lower panel: ISGRI light curve
shows a constant flux level (revolutions 624 and 625 are co-added due to the low exposure of the
former revolution). 
\label{fig:lc}}
\end{figure}

\subsection{Variability analysis}
Blazars are strongly variable objects on timescales as short as minutes 
to hours (e.g., \citealt{fossati08}). The long exposures used with 
IBIS/ISGRI may therefore be affected by variations in flux and  
spectrum. Thus, we analyzed the available multiwavelength data set for 
indications of variability.

The non-simultaneous (to IBIS/ISGRI, {\it Swift}/XRT and {\it Swift}/UVOT; see Table~\ref{tab:uvot}) 
Palomar 60- inch observations show a very low variability in brightness of the source of $10$\,\% at 
a significance of $\sim$ 3$\sigma$.

The UVOT observations show marginal brightness variations by about 
10\,\% on a daily timescale and 20\,\% on inter-day timescales (significant 
at $\sim$ 6$\sigma$).

The all-sky monitor (ASM; \citealt{levine96}) on the {\it RXTE)} provides regular 
monitoring of 1ES~1959+650 in the 1.5 -- 10~keV band
\footnote{http://xte.mit.edu/asmlc/ASM.html}.
The dwell-by-dwell 90\,s exposure light curve does not show flaring activity during the 
{\it INTEGRAL}/IBIS observations (Figure~\ref{fig:lc}, upper panel). The one-day average light curve
excludes the presence of any flare, and it shows flux variability ($<20$\,\%), well within the errors.

The 20-40\,keV IBIS/ISGRI light curve binned per revolution is shown
in Figure~\ref{fig:lc} - lower panel (revolutions 624 and 625 are co-added).
The low detection significance of the source in each
individual revolution ($\sim 5\sigma$) does not allow a proper
light-curve fitting. If binned to SCW level (time binning of $\sim$1 hr), the
light curve does not allow to rule out some marginally significant variability
due to the low signal-to-noise ratio (S/N). However, we can exclude any flaring 
activity in the 20-40\,keV band.

\subsection{{\it Swift}/XRT and IBIS/ISGRI spectra}
The fit results for the single {\it Swift}/XRT and averaged IBIS/ISGRI spectrum are shown 
in Table~\ref{tab:fit}. The spectra were fitted using XSPEC 12 and 
the latest available response matrices for calibration. IBIS/ISGRI detected 1ES~1959+650 
with a flux of $\sim$ 1.5 $\times$ 10$^{-11}$ erg cm$^{-2}$ s$^{-1}$ in the 20 -- 60 keV range. 
The low Galactic N$_{H}$- value (1.0 $\times$ 10$^{21}$ cm$^{-2}$; \citealt{dickey90}) 
does not affect the IBIS/ISGRI energy range, and we do not see evidence for further 
absorption or deviations from a single power law.

The best-fit result ($\chi^{2}_{red}$ $\sim$ 0.9) for the {\it Swift}/XRT spectrum 
is given by an absorbed power-law model, with absorption parameter free to vary. The
derived N$_{H}$-value (1.7$^{2.1}_{1.4}$ $\times$ 10$^{21}$ cm$^{-2}$) is a factor 1.7
larger (at a significance of $\sim$ 2$\sigma$) than the Galactic one.

We do not see significant variability
neither on long timescales nor on short timescales in the hard X-ray
(20 -- 40~keV) and in the X-ray (1.5 -- 10~keV) bands during the multifrequency campaign
(see Figure~\ref{fig:lc}). Therefore, we combine the relatively short (simultaneous to IBIS/ISGRI) 
{\it Swift}/XRT observation (0.6 -- 6~keV) with the averaged IBIS/ISGRI spectrum.
We model the data best with a broken power-law model with the absorption parameter fixed
to the Galactic value. Furthermore, we have applied a log-parabolic model:
\begin{equation}
F(E) = K~ (E/E_1)^{-(a + b~log(E/E_1))}.  
\end{equation}
This curved model was first proposed by \citet{landau86} to describe the synchrotron
component of BL Lac objects. \citet{massaro04} used it to describe the 
synchrotron X-ray component of the spectrum of TeV BL Lac objects. 
The model is explained in physical terms by means of radiative cooling processes 
(via synchrotron and inverse-Compton) of the high-energy electron population injected 
with power-law slope. The log-parabolic law is a rather simple analytical formula related
to physical parameters of the source. It is applied to synchrotron broadband spectra since it better
describes the spectra compared to power laws with exponential cutoff. 
The photon index $a$ is considered at energy $E_1$, and the parameter $b$ describes the 
curvature of the spectrum.  
The fit result is reported in Table ~\ref{tab:fit}.

\section{The spectral energy distribution}

\begin{table*}
\begin{minipage}[]{\textwidth}
\begin{center}
\centering
\caption{List of parameters used to construct the theoretical SEDs.}\label{tab:parameters}
\begin{tabular}{cccccccccccc}
\hline
\hline
$Model$ & $R$ & $\eta_{\rm esc}$ & $B$ & D & $\gamma_{\rm min}$ & $\gamma_{\rm max}$ & $q$ & $\gamma_{\rm pmin}$ & $\gamma_{\rm pmax}$ & $q_{\rm p}$ & $L_{\rm p}$\\
(1) & (2) & (3) & (4) & (5) & (6) & (7) & (8) & (9) & (10) & (11) & (12)\\
\hline   
leptonic       & 0.5 & 10 & 14  & 19 & 1$\times$10$^{3}$ & 6$\times$10$^{4}$ & 1.85 & - & - & - & -\\ 
lepto-hadronic & 2 & 5 & 20 & 19 & 8$\times$10$^{2}$ & 4.5$\times$10$^{4}$ & 1.9 & 1$\times$10$^{3}$ & 1.2$\times$10$^{9}$ & 1.9 & 3.5\\  
\hline
\hline 
\end{tabular}
\end{center}
\end{minipage}
Explanation of columns:
(1) = SED model;
(2) = radius of emitting region in units of $10^{14}$~cm;
(3) = escape time parameter: $t_{esc} = \eta \times R/c$;
(4) = magnetic field in Gauss;
(5) = Doppler factor;
(6) and (7) = minimum and maximum random Lorentz factors of the injected electrons;
(8) = slope of the injected electron distribution;
(9) and (10) = minimum and maximum random Lorentz factors of the injected protons;
(11) = slope of the injected proton distribution; 
(12) = kinetic power of relativistic protons in units of 10$^{46}$ erg~s$^{-1}$.
\end{table*}

\begin{figure}
\epsscale{1.00}
\plotone{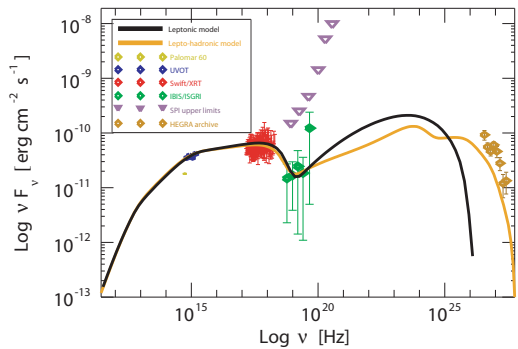}
\caption{Observed SED of 1ES~1959+650 in 2007 November and December. 
The model fits to the SED (see the text) are overplotted (black solid and orange solid curves
are the leptonic and lepto-hadronic models respectively).
$\gamma\gamma$ absorption by the extragalactic background light (EBL) is accounted. 
\label{fig:fig02}}
\end{figure}

In order to construct a multiwavelength spectrum of the source we use the data sampled 
during the simultaneous multifrequency campaign in 2007. At TeV energies we show non 
simultaneous archive HEGRA IACT data \citep{aharonian03}, when the source showed a major 
outburst (in 2002). Such activity at TeV energies was not reported during our multifrequency 
campaign in 2007. Therefore these data are not used to constrain our SED model.

Given the lack of variability on short and long timescales, we used the average values in each 
energy band to construct the SED. The SED of 1ES~1959+650 shows the typical two-component 
structure of an HBL (Figure \ref{fig:fig02}). 
The synchrotron peak can be constrained by the XRT spectrum ($a$ $\sim$ 2,
see Table~\ref{tab:fit}), whereas the hard X-ray--soft $\gamma$-ray data constrain the
onset of the high-energy component. Indeed, the log-parabolic fit model for the synchrotron 
emission is not able to reproduce the jointly fitted XRT and IBIS/ISGRI data (see fit result 
in Table~\ref{tab:fit}). This, in turn, suggests that the IBIS/ISGRI spectrum represents 
the onset of the high-energy component.
The high-energy peak is very poorly constrained due to the lack of simultaneous $\gamma$-ray 
data. 
During its first year of survey, the {\it Fermi} Large Area Telescope (LAT) detects the source
in quiescence state \citep{abdo10}. Therefore the power output measured by LAT is not comparable
to the power output in the same energy range inferred with our model that refers to an active 
state of the source. The LAT measurement is a factor $\sim$6 lower. The {\it Fermi}-LAT 
observations are taken $\sim$1 year after our multifrequency campaign of 2007.

The data are modeled by a pure leptonic SSC model (see Figure~\ref{fig:fig02} black solid line), 
using the equilibrium version of the code of \citet{boettcher02b}, as described in more detail 
in \citep{acciari09}. The geometry of the emitting region is a spherical volume $V'_b$ 
of radius $R_{\rm b}$ in the comoving frame. It moves with respect to the observer with 
a bulk Lorentz factor $\Gamma$ (speed $\beta_{\Gamma}$c) at an angle $\theta_{\rm obs}$, 
resulting in relativistic beaming determined by the Doppler factor. Ultrarelativistic leptons are 
injected into the emission region with a power-law distribution (in the comoving frame):
\begin{equation}
Q_e^{\rm inj} (\gamma; t) = Q_0^{\rm inj} (t) \, \gamma^{-q} \; \; [cm^{-3} s^{-1}] \; \; 
{\rm for} \;\; \gamma_1 \le \gamma \le \gamma_2
\label{Qe}
\end{equation}
where the normalization is determined by the injection power $L_{\rm inj}$.
The code finds a self-consistent equilibrium between particle injection,
radiative cooling due to synchrotron and synchrotron self-Compton (SSC)
losses and particle escape on a timescale $t_{\rm esc} = \eta_{\rm esc} R/c$.
The effect of $\gamma\gamma$ absorption by the extragalactic background light
(EBL) is taken into account using the model of \citet{finke10}. 
The fit is constrained by the synchrotron component from the optical to
X-rays as well as the onset of the SSC component at hard X-rays -- soft $\gamma$-rays.
The hard synchrotron spectrum as well as the unusually high level of the low-frequency 
end of the SSC component (see Figure \ref{fig:fig02}) requires the choice of a very hard 
injection spectrum with a slope of $q = 1.85$. The optical data point 
from the Palomar observations were taken up to 10 days after the X-ray and UV pointings 
with XRT and UVOT respectively. They cannot be reconciled with our model SED. This is
not surprising, as variability by a factor of $\sim 2$ on timescales of weeks is not
uncommon in this object \citep{villata00}. 
The parameters used for the fit shown in Figure \ref{fig:fig02} are listed in Table 
\ref{tab:parameters}. The equilibrium particle distribution found by the code 
corresponds to a kinetic power in relativistic electrons of $L_e = 8.5 \times 
10^{42}$ erg~s$^{-1}$, while the magnetic field of $B = 14$~G yields to a power in 
Poynting flux of $L_B = 7.4 \times 10^{41}$ erg~s$^{-1}$. Hence, the magnetic field 
energy density is a factor $\epsilon_B \equiv L_B/L_e = 0.09$ below equipartition.
The hard injection index of $q = 1.85$ (constrained by the mere synchrotron component) 
is inconsistent with standard first-order
Fermi acceleration at relativistic shocks, which predicts an index of $q \sim 2.2$
-- 2.3 \citep{achterberg01,ellison04}. This might indicate a substantial 
contribution to particle acceleration from second-order Fermi acceleration
\citep{virtanen05,stecker07}. A similar conclusion was also reached when modeling 
the very hard X-ray and {\it Fermi} $\gamma$-ray spectrum of the HBL RGB~J0710+591, 
recently detected at very-high-energy (VHE) $\gamma$-rays by VERITAS. 

As an alternative to the pure leptonic model, we have applied a semi-analytical
lepto-hadronic model shown by the orange solid line in Figure \ref{fig:fig02}.
This model assumes, in addition to a leptonic component similar
to the one used for the leptonic model described above, a
power-law population of relativistic protons extending out
to energies beyond the threshold for p$\gamma$ pion production
on the electron-synchrotron radiation field. The production
rates of final decay products (electrons, positrons,
$\pi^0$ decay photons, and neutrinos) are calculated using
the analytical templates of \citet{kelner08}. Synchrotron emission 
of secondaries is calculated using a $j_{\nu} (\gamma) 
\propto \nu^{1/3} e^{-\nu/\nu_0 (\gamma)}$ approximation.
The $\pi^0$ decay photons as well as synchrotron emission
from the first-generation pairs from charged pion decay
are produced predominantly at $\gg$~TeV energies, at which
the emission region is highly opaque to $\gamma\gamma$
absorption. Therefore, the radiative power at those energies
is redistributed to lower frequencies through electromagnetic
cascades. We employ a semi-analytical treatment of the cascading
process as described in \citet{boettcher10}.
The inferred parameters are reported in Tabel \ref{tab:parameters}, 
and they are in good agreement with the leptonic model.
\section{Conclusions}
We have performed a multiwavelength campaign on 1ES~1959+650 in 2007. This is the 
first and only detection by IBIS/ISGRI of the source. The source was monitored nearly 
simultaneously from optical to hard X-ray energies. In the optical to UV band the most 
evident flux variations are of the order of 20\,\% on inter-day timescales. 
At hard X-rays the source flux stayed constant. The compiled SED allowed us to derive 
the physical parameters of the source. The SED clearly shows the simultaneous detection 
of both the synchrotron and the rise of the high-energy (inverse-Compton) emission components.
This is the first time that the transition region between the synchrotron and inverse-Compton
components is detected for 1ES~1959+650, and has so far been measured only for a few
other BL Lac objects of intermediate type. The best example is ON 231 \citep{tagliaferri00} 
which is an Intermediate BL Lac object (IBL). These objects are characterized by SEDs 
peaking at frequencies intermediate to LBL and HBL. Other examples, not as clear as the 
previous one, are PKS 2155-304 \citep{kubo98, foschini08}, another IBL, and 
S5 0716+714 \citep{foschini06, giommi99}, an LBL. \citet{foschini08} suggest 
that the source is in a continuous high active state and only seldom lowers its 
activity shifting the position of the peak of the synchrotron emission to lower 
frequencies. 
For HBL objects, the synchrotron peak is located usually at soft X-ray 
energies and exceptionally at hard X-ray energies, thus making 
the descending branch of the synchrotron component and the ascending 
branch of the inverse-Compton component difficult or impossible 
to measure. This interesting hard X-ray regime (20~keV -- 1~MeV) is important since the two 
emission components are competing: higher fluxes translate into harder spectra
indicating changes in the injected particle population. In turn this gives clues 
on the jet physics and its composition. Monitoring the hard X-ray to $\gamma$-ray
domain allows to trace the evolution of the emission mechanism. This can be performed
by instruments with higher sensitivity as could be GRIPS \citep{greiner09}, a future
$\gamma$-ray mission proposed to the European Space Agency.

We reproduced the observed SED with a simple one-zone, leptonic SSC model and with a
lepto-hadronic model. Both models required a very hard electron 
injection spectrum with an index of $q = 1.85$ and $q = 1.9$ for the leptonic
and lepto-hadronic models respectively. This requirement might 
indicate the importance of second-order Fermi acceleration mechanisms in the
energization of ultrarelativistic particles in the jet of 1ES~1959+650.

\acknowledgments
We thank the {\it INTEGRAL} and the {\it Swift} team for the observations and the support. 
The anonymous referee is acknowledged for her/his helpful comments which improved
the manuscript. E.B. acknowledges the ISDC for the warm hospitality during the NEP 
monitoring. The work of M.B. was partially supported by NASA through {\it INTEGRAL} Guest 
Observer Grant NNX09A171G and Fermi Guest Investigator Grant NNX09AT82G.
A.R. is grateful for support through NASA grant NNX08AY13G. X.-L.Z. acknowledges financial 
support by DLR FKZ50OG0502.


\begin{thebibliography}{50}


\bibitem[{{Abdo} {et~al.}(2010){Abdo}, {Ackermann}, {Ajello}, {Allafort},
  {Antolini}, {Atwood}, {Axelsson}, {Baldini}, {Ballet}, {Barbiellini},
  {Bastieri}, {Baughman}, {Bechtol}, {Bellazzini}, {Berenji}, {Blandford},
  {Bloom}, {Bogart}, {Bonamente}, {Borgland}, {Bouvier}, {Bregeon}, {Brez},
  {Brigida}, {Bruel}, {Buehler}, {Burnett}, {Buson}, {Caliandro}, {Cameron},
  {Cannon}, {Caraveo}, {Carrigan}, {Casandjian}, {Cavazzuti}, {Cecchi}, {{\c
  C}elik}, {Celotti}, {Charles}, {Chekhtman}, {Chen}, {Cheung}, {Chiang},
  {Ciprini}, {Claus}, {Cohen-Tanugi}, {Conrad}, {Costamante}, {Cotter},
  {Cutini}, {D'Elia}, {Dermer}, {de Angelis}, {de Palma}, {De Rosa}, {Digel},
  {Silva}, {Drell}, {Dubois}, {Dumora}, {Escande}, {Farnier}, {Favuzzi},
  {Fegan}, {Ferrara}, {Focke}, {Fortin}, {Frailis}, {Fukazawa}, {Funk},
  {Fusco}, {Gargano}, {Gasparrini}, {Gehrels}, {Germani}, {Giebels},
  {Giglietto}, {Giommi}, {Giordano}, {Giroletti}, {Glanzman}, {Godfrey},
  {Grandi}, {Grenier}, {Grondin}, {Grove}, {Guiriec}, {Hadasch}, {Harding},
  {Hayashida}, {Hays}, {Healey}, {Hill}, {Horan}, {Hughes}, {Iafrate}, {Itoh},
  {J{\'o}hannesson}, {Johnson}, {Johnson}, {Johnson}, {Johnson}, {Kamae},
  {Katagiri}, {Kataoka}, {Kawai}, {Kerr}, {Kn{\"o}dlseder}, {Kuss}, {Lande},
  {Latronico}, {Lavalley}, {Lemoine-Goumard}, {Llena Garde}, {Longo},
  {Loparco}, {Lott}, {Lovellette}, {Lubrano}, {Madejski}, {Makeev}, {Malaguti},
  {Massaro}, {Mazziotta}, {McConville}, {McEnery}, {McGlynn}, {Michelson},
  {Mitthumsiri}, {Mizuno}, {Moiseev}, {Monte}, {Monzani}, {Morselli},
  {Moskalenko}, {Murgia}, {Nolan}, {Norris}, {Nuss}, {Ohno}, {Ohsugi},
  {Omodei}, {Orlando}, {Ormes}, {Ozaki}, {Paneque}, {Panetta}, {Parent},
  {Pelassa}, {Pepe}, {Pesce-Rollins}, {Piranomonte}, {Piron}, {Porter},
  {Rain{\`o}}, {Rando}, {Razzano}, {Reimer}, {Reimer}, {Reposeur}, {Ripken},
  {Ritz}, {Rodriguez}, {Romani}, {Roth}, {Ryde}, {Sadrozinski}, {Sanchez},
  {Sander}, {Saz Parkinson}, {Scargle}, {Sgr{\`o}}, {Shaw}, {Siskind}, {Smith},
  {Spandre}, {Spinelli}, {Starck}, {Stawarz}, {Strickman}, {Suson}, {Tajima},
  {Takahashi}, {Takahashi}, {Tanaka}, {Taylor}, {Thayer}, {Thayer}, {Thompson},
  {Tibaldo}, {Torres}, {Tosti}, {Tramacere}, {Ubertini}, {Uchiyama}, {Usher},
  {Vasileiou}, {Vilchez}, {Villata}, {Vitale}, {Waite}, {Wallace}, {Wang},
  {Winer}, {Wood}, {Yang}, {Ylinen}, \& {Ziegler}}]{abdo10}
{Abdo}, A.~A., {Ackermann}, M., {Ajello}, M., {et~al.} 2010, \apj, 715, 429

\bibitem[{{Acciari} {et~al.}(2009){Acciari}, {Aliu}, {Aune}, {Beilicke},
  {Benbow}, {B{\"o}ttcher}, {Boltuch}, {Buckley}, {Bradbury}, {Bugaev},
  {Byrum}, {Cannon}, {Cesarini}, {Ciupik}, {Cogan}, {Cui}, {Dickherber},
  {Duke}, {Falcone}, {Finley}, {Fortin}, {Fortson}, {Furniss}, {Galante},
  {Gall}, {Gibbs}, {Gillanders}, {Grube}, {Guenette}, {Gyuk}, {Hanna},
  {Holder}, {Hui}, {Humensky}, {Kaaret}, {Karlsson}, {Kertzman}, {Kieda},
  {Konopelko}, {Krawczynski}, {Krennrich}, {Lang}, {Le Bohec}, {Maier},
  {McArthur}, {McCann}, {McCutcheon}, {Millis}, {Moriarty}, {Ong}, {Otte},
  {Pandel}, {Perkins}, {Pichel}, {Pohl}, {Quinn}, {Ragan}, {Reyes}, {Reynolds},
  {Roache}, {Rose}, {Sembroski}, {Smith}, {Steele}, {Theiling}, {Thibadeau},
  {Varlotta}, {Vassiliev}, {Vincent}, {Wakely}, {Ward}, {Weekes}, {Weinstein},
  {Weisgarber}, {Williams}, {Wissel}, {Wood}, {Pian}, {Vercellone},
  {Donnarumma}, {D'Ammando}, {Bulgarelli}, {Chen}, {Giuliani}, {Longo},
  {Pacciani}, {Pucella}, {Vittorini}, {Tavani}, {Argan}, {Barbiellini},
  {Caraveo}, {Cattaneo}, {Cocco}, {Costa}, {Del Monte}, {De Paris}, {Di Cocco},
  {Evangelista}, {Feroci}, {Fiorini}, {Froysland}, {Frutti}, {Fuschino},
  {Galli}, {Gianotti}, {Labanti}, {Lapshov}, {Lazzarotto}, {Lipari},
  {Marisaldi}, {Mastropietro}, {Mereghetti}, {Morelli}, {Morselli},
  {Pellizzoni}, {Perotti}, {Piano}, {Picozza}, {Pilia}, {Porrovecchio},
  {Prest}, {Rapisarda}, {Rappoldi}, {Rubini}, {Sabatini}, {Soffitta},
  {Trifoglio}, {Trois}, {Vallazza}, {Zambra}, {Zanello}, {Pittori},
  {Santolamazza}, {Verrecchia}, {Giommi}, {Colafrancesco}, {Salotti},
  {Villata}, {Raiteri}, {Aller}, {Aller}, {Arkharov}, {Efimova}, {Larionov},
  {Leto}, {Ligustri}, {Lindfors}, {Pasanen}, {Kurtanidze}, {Tetradze},
  {Lahteenmaki}, {Kotiranta}, {Cucchiara}, {Romano}, {Nesci}, {Pursimo},
  {Heidt}, {Benitez}, {Hiriart}, {Nilsson}, {Berdyugin}, {Mujica}, {Dultzin},
  {Lopez}, {Mommert}, {Sorcia}, \& {de la Calle Perez}}]{acciari09}
{Acciari}, V.~A., {Aliu}, E., {Aune}, T., {et~al.} 2009, \apj, 707, 612

\bibitem[{{Achterberg} {et~al.}(2001){Achterberg}, {Gallant}, {Kirk}, \&
  {Guthmann}}]{achterberg01}
{Achterberg}, A., {Gallant}, Y.~A., {Kirk}, J.~G., \& {Guthmann}, A.~W. 2001,
  \mnras, 328, 393

\bibitem[{{Aharonian} {et~al.}(2003){Aharonian}, {Akhperjanian}, {Beilicke},
  {Bernloehr}, {Boerst}, {Bojahr}, {Bolz}, {Coarasa}, {Contreras}, {Cortina},
  \& {Denninghoff}}]{aharonian03}
{Aharonian}, F., {Akhperjanian}, A., {Beilicke}, M., {et~al.} 2003, VizieR
  Online Data Catalog, 340, 69009


\bibitem[{{Antonucci}(1993)}]{antonucci93}
{Antonucci}, R. 1993, \araa, 31, 473

\bibitem[{{Beckmann} {et~al.}(2002){Beckmann}, {Wolter}, {Celotti},
  {Costamante}, {Ghisellini}, {Maccacaro}, \& {Tagliaferri}}]{beckmann02}
{Beckmann}, V., {Wolter}, A., {Celotti}, A., {et~al.} 2002, \aap, 383, 410

\bibitem[{{Bottacini} {et~al.}(2007){Bottacini}, {Beckmann}, {Ishibashi},
  {Ajello}, \& {Greiner}}]{bottacini07}
{Bottacini}, E., {Beckmann}, V., {Ishibashi}, W., {Ajello}, M., \& {Greiner},
  J. 2007, The Astronomer's Telegram, 1315, 1

\bibitem[{{B{\"o}ttcher}(2005)}]{boettcher05}
{B{\"o}ttcher}, M. 2005, \apj, 621, 176

\bibitem[{{B{\"o}ttcher}(2007)}]{boettcher07}
{B{\"o}ttcher}, M. 2007, \apss, 309, 95

\bibitem[{{B{\"o}ttcher}(2010)}]{boettcher10}
{B{\"o}ttcher}, M. 2010, Proc. of Fermi Meets Jansky, MPIfR Bonn, Eds.: 
T. Savolainen, E. Ros, R. W. Porcas \& J. A. Zensus; p. 41

\bibitem[{{B{\"o}ttcher} \& {Chiang}(2002)}]{boettcher02b}
{B{\"o}ttcher}, M. \& {Chiang}, J. 2002, \apj, 581, 127

\bibitem[{{Burrows} {et~al.}(2005){Burrows}, {Hill}, {Nousek}, {Kennea},
  {Wells}, {Osborne}, {Abbey}, {Beardmore}, {Mukerjee}, {Short}, {Chincarini},
  {Campana}, {Citterio}, {Moretti}, {Pagani}, {Tagliaferri}, {Giommi},
  {Capalbi}, {Tamburelli}, {Angelini}, {Cusumano}, {Br{\"a}uninger}, {Burkert},
  \& {Hartner}}]{burrows05}
{Burrows}, D.~N., {Hill}, J.~E., {Nousek}, J.~A., {et~al.} 2005, Space Science
  Reviews, 120, 165

\bibitem[{{Cardelli} {et~al.}(1989){Cardelli}, {Clayton}, \&
  {Mathis}}]{cardelli89}
{Cardelli}, J.~A., {Clayton}, G.~C., \& {Mathis}, J.~S. 1989, \apj, 345, 245

\bibitem[{{Cenko} {et~al.}(2006){Cenko}, {Fox}, {Moon}, {Harrison}, {Kulkarni},
  {Henning}, {Guzman}, {Bonati}, {Smith}, {Thicksten}, {Doyle}, {Petrie},
  {Gal-Yam}, {Soderberg}, {Anagnostou}, \& {Laity}}]{cenko06}
{Cenko}, S.~B., {Fox}, D.~B., {Moon}, D.-S., {et~al.} 2006, \pasp, 118, 1396

\bibitem[{{Courvoisier} {et~al.}(2003){Courvoisier}, {Walter}, {Beckmann},
  {Dean}, {Dubath}, {Hudec}, {Kretschmar}, {Mereghetti}, {Montmerle},
  {Mowlavi}, {Paltani}, {Preite Martinez}, {Produit}, {Staubert}, {Strong},
  {Swings}, {Westergaard}, {White}, {Winkler}, \& {Zdziarski}}]{courvoisier03}
{Courvoisier}, T.~J.-L., {Walter}, R., {Beckmann}, V., {et~al.} 2003, \aap,
  411, L53

\bibitem[{{Dermer} \& {Schlickeiser}(1993)}]{dermer93}
{Dermer}, C.~D. \& {Schlickeiser}, R. 1993, \apj, 416, 458

\bibitem[{{Dickey} \& {Lockman}(1990)}]{dickey90}
{Dickey}, J.~M. \& {Lockman}, F.~J. 1990, \araa, 28, 215

\bibitem[{{Ellison} \& {Double}(2004)}]{ellison04}
{Ellison}, D.~C. \& {Double}, G.~P. 2004, Astroparticle Physics, 22, 323

\bibitem[{{Elvis} {et~al.}(1992){Elvis}, {Plummer}, {Schachter}, \&
  {Fabbiano}}]{elvis92}
{Elvis}, M., {Plummer}, D., {Schachter}, J., \& {Fabbiano}, G. 1992, \apjs, 80,
  257

\bibitem[{{Finke} {et~al.}(2010){Finke}, {Razzaque}, \& {Dermer}}]{finke10}
{Finke}, J.~D., {Razzaque}, S., \& {Dermer}, C.~D. 2010, \apj, 712, 238

\bibitem[{{Foschini} {et~al.}(2006){Foschini}, {Tagliaferri}, {Pian},
  {Ghisellini}, {Treves}, {Maraschi}, {Tavecchio}, {Di Cocco}, \&
  {Rosen}}]{foschini06}
{Foschini}, L., {Tagliaferri}, G., {Pian}, E., {et~al.} 2006, \aap, 455, 871

\bibitem[{{Foschini} {et~al.}(2008){Foschini}, {Treves}, {Tavecchio},
  {Impiombato}, {Ghisellini}, {Covino}, {Tosti}, {Gliozzi}, {Bianchin}, {Di
  Cocco}, {Malaguti}, {Maraschi}, {Pian}, {Raiteri}, {Sambruna}, {Tagliaferri},
  \& {Villata}}]{foschini08}
{Foschini}, L., {Treves}, A., {Tavecchio}, F., {et~al.} 2008, \aap, 484, L35

\bibitem[{{Fossati} {et~al.}(2008){Fossati}, {Buckley}, {Bond}, {Bradbury},
  {Carter-Lewis}, {Chow}, {Cui}, {Falcone}, {Finley}, {Gaidos}, {Grube},
  {Holder}, {Horan}, {Horns}, {Jordan}, {Kieda}, {Kildea}, {Krawczynski},
  {Krennrich}, {Lang}, {LeBohec}, {Lee}, {Moriarty}, {Ong}, {Petry}, {Quinn},
  {Sembroski}, {Wakely}, \& {Weekes}}]{fossati08}
{Fossati}, G., {Buckley}, J.~H., {Bond}, I.~H., {et~al.} 2008, \apj, 677, 906


\bibitem[{{Giommi} {et~al.}(1999){Giommi}, {Massaro}, {Chiappetti}, {Ferrara},
  {Ghisellini}, {Jang}, {Maesano}, {Miller}, {Montagni}, {Nesci}, {Padovani},
  {Perlman}, {Raiteri}, {Sclavi}, {Tagliaferri}, {Tosti}, \&
  {Villata}}]{giommi99}
{Giommi}, P., {Massaro}, E., {Chiappetti}, L., {et~al.} 1999, \aap, 351, 59

\bibitem[{{Greiner} {et~al.}(2009){Greiner}, {Iyudin}, \& {Kanbach}}]{greiner09}
{Greiner}, J., {Iyudin}, A., \& {Kanbach}, G. 2009, Exp. Astr., 23, 91

\bibitem[{{Hartman} {et~al.}(1999){Hartman}, {Bertsch}, {Bloom}, {Chen},
  {Deines-Jones}, {Esposito}, {Fichtel}, {Friedlander}, {Hunter}, {McDonald},
  {Sreekumar}, {Thompson}, {Jones}, {Lin}, {Michelson}, {Nolan}, {Tompkins},
  {Kanbach}, {Mayer-Hasselwander}, {M{\"u}cke}, {Pohl}, {Reimer}, {Kniffen},
  {Schneid}, {von Montigny}, {Mukherjee}, \& {Dingus}}]{hartman99}
{Hartman}, R.~C., {Bertsch}, D.~L., {Bloom}, S.~D., {et~al.} 1999, \apjs, 123,
  79
  
\bibitem[{{Kelner} \& {Aharonian}(2008)}]{kelner08}
  {Kelner}, S.~R. \& {Aharonian}, F.~A. 2008, \prd, 78, 034013

\bibitem[{{Krawczynski} {et~al.}(2004){Krawczynski}, {Hughes}, {Horan},
  {Aharonian}, {Aller}, {Aller}, {Boltwood}, {Buckley}, {Coppi}, {Fossati},
  {G{\"o}tting}, {Holder}, {Horns}, {Kurtanidze}, {Marscher}, {Nikolashvili},
  {Remillard}, {Sadun}, \& {Schr{\"o}der}}]{krawczynski04}
{Krawczynski}, H., {Hughes}, S.~B., {Horan}, D., {et~al.} 2004, \apj, 601, 151

\bibitem[{{Kubo} {et~al.}(1998){Kubo}, {Takahashi}, {Madejski}, {Tashiro},
  {Makino}, {Inoue}, \& {Takahara}}]{kubo98}
{Kubo}, H., {Takahashi}, T., {Madejski}, G., {et~al.} 1998, \apj, 504, 693

\bibitem[{{Landau} {et~al.}(1986){Landau}, {Golisch}, {Jones}, {Jones},
  {Pedelty}, {Rudnick}, {Sitko}, {Kenney}, {Roellig}, {Salonen}, {Urpo},
  {Schmidt}, {Neugebauer}, {Matthews}, {Elias}, {Impey}, {Clegg}, \&
  {Harris}}]{landau86}
{Landau}, R., {Golisch}, B., {Jones}, T.~J., {et~al.} 1986, \apj, 308, 78

\bibitem[{{Lebrun} {et~al.}(2003){Lebrun}, {Leray}, {Lavocat}, {Cr{\'e}tolle},
  {Arqu{\`e}s}, {Blondel}, {Bonnin}, {Bou{\`e}re}, {Cara}, {Chaleil}, {Daly},
  {Desages}, {Dzitko}, {Horeau}, {Laurent}, {Limousin}, {Mathy}, {Mauguen},
  {Meignier}, {Molini{\'e}}, {Poindron}, {Rouger}, {Sauvageon}, \&
  {Tourrette}}]{lebrun03}
{Lebrun}, F., {Leray}, J.~P., {Lavocat}, P., {et~al.} 2003, \aap, 411, L141

\bibitem[{{Levine} {et~al.}(1996){Levine}, {Bradt}, {Cui}, {Jernigan},
  {Morgan}, {Remillard}, {Shirey}, \& {Smith}}]{levine96}
{Levine}, A.~M., {Bradt}, H., {Cui}, W., {et~al.} 1996, \apjl, 469, L33+

\bibitem[{{Lund} {et~al.}(2003){Lund}, {Budtz-J{\o}rgensen}, {Westergaard},
  {Brandt}, {Rasmussen}, {Hornstrup}, {Oxborrow}, {Chenevez}, {Jensen},
  {Laursen}, {Andersen}, {Mogensen}, {Rasmussen}, {Om{\o}}, {Pedersen},
  {Polny}, {Andersson}, {Andersson}, {K{\"a}m{\"a}r{\"a}inen}, {Vilhu},
  {Huovelin}, {Maisala}, {Morawski}, {Juchnikowski}, {Costa}, {Feroci},
  {Rubini}, {Rapisarda}, {Morelli}, {Carassiti}, {Frontera}, {Pelliciari},
  {Loffredo}, {Mart{\'{\i}}nez N{\'u}{\~n}ez}, {Reglero}, {Velasco}, {Larsson},
  {Svensson}, {Zdziarski}, {Castro-Tirado}, {Attina}, {Goria}, {Giulianelli},
  {Cordero}, {Rezazad}, {Schmidt}, {Carli}, {Gomez}, {Jensen}, {Sarri},
  {Tiemon}, {Orr}, {Much}, {Kretschmar}, \& {Schnopper}}]{lund03}
{Lund}, N., {Budtz-J{\o}rgensen}, C., {Westergaard}, N.~J., {et~al.} 2003,
  \aap, 411, L231

\bibitem[{{Mannheim}(1993)}]{mannheim93}
{Mannheim}, K. 1993, \aap, 269, 67

\bibitem[{{Mas-Hesse} {et~al.}(2003){Mas-Hesse}, {Gim{\'e}nez}, {Culhane},
  {Jamar}, {McBreen}, {Torra}, {Hudec}, {Fabregat}, {Meurs}, {Swings},
  {Alcacera}, {Balado}, {Beiztegui}, {Belenguer}, {Bradley}, {Caballero},
  {Cabo}, {Defise}, {D{\'{\i}}az}, {Domingo}, {Figueras}, {Figueroa}, {Hanlon},
  {Hroch}, {Hudcova}, {Garc{\'{\i}}a}, {Jordan}, {Jordi}, {Kretschmar},
  {Laviada}, {March}, {Mart{\'{\i}}n}, {Mazy}, {Men{\'e}ndez}, {Mi}, {de
  Miguel}, {Mu{\~n}oz}, {Nolan}, {Olmedo}, {Plesseria}, {Polcar}, {Reina},
  {Renotte}, {Rochus}, {S{\'a}nchez}, {San Mart{\'{\i}}n}, {Smith}, {Soldan},
  {Thomas}, {Tim{\'o}n}, \& {Walton}}]{mas-hesse03}
{Mas-Hesse}, J.~M., {Gim{\'e}nez}, A., {Culhane}, J.~L., {et~al.} 2003, \aap,
  411, L261

\bibitem[{{Massaro} {et~al.}(2004){Massaro}, {Perri}, {Giommi}, \&
  {Nesci}}]{massaro04}
{Massaro}, E., {Perri}, M., {Giommi}, P., \& {Nesci}, R. 2004, \aap, 413, 489

\bibitem[{{M{\"u}cke} {et~al.}(2003){M{\"u}cke}, {Protheroe}, {Engel},
  {Rachen}, \& {Stanev}}]{muecke03}
{M{\"u}cke}, A., {Protheroe}, R.~J., {Engel}, R., {Rachen}, J.~P., \& {Stanev},
  T. 2003, Astroparticle Physics, 18, 593

\bibitem[{{Petry} {et~al.}(2009){Petry}, {Beckmann}, {Halloin}, \&
  {Strong}}]{petry09}
{Petry}, D., {Beckmann}, V., {Halloin}, H., \& {Strong}, A. 2009, \aap, 507,
  549

\bibitem[{{Rees}(1966)}]{rees66}
{Rees}, M.~J. 1966, \nat, 211, 468

\bibitem[{{Roming} {et~al.}(2005){Roming}, {Kennedy}, {Mason}, {Nousek}, {Ahr},
  {Bingham}, {Broos}, {Carter}, {Hancock}, {Huckle}, {Hunsberger}, {Kawakami},
  {Killough}, {Koch}, {McLelland}, {Smith}, {Smith}, {Soto}, {Boyd},
  {Breeveld}, {Holland}, {Ivanushkina}, {Pryzby}, {Still}, \&
  {Stock}}]{roming05}
{Roming}, P.~W.~A., {Kennedy}, T.~E., {Mason}, K.~O., {et~al.} 2005, Space
  Science Reviews, 120, 95

\bibitem[{{Schachter} {et~al.}(1993){Schachter}, {Stocke}, {Perlman}, {Elvis},
  {Remillard}, {Granados}, {Luu}, {Huchra}, {Humphreys}, {Urry}, \&
  {Wallin}}]{schachter93}
{Schachter}, J.~F., {Stocke}, J.~T., {Perlman}, E., {et~al.} 1993, \apj, 412,
  541

\bibitem[{{Schlegel} {et~al.}(1998){Schlegel}, {Finkbeiner}, \&
  {Davis}}]{schlegel98}
{Schlegel}, D.~J., {Finkbeiner}, D.~P., \& {Davis}, M. 1998, \apj, 500, 525

\bibitem[{{Sikora} {et~al.}(1994){Sikora}, {Begelman}, \& {Rees}}]{sikora94}
{Sikora}, M., {Begelman}, M.~C., \& {Rees}, M.~J. 1994, \apj, 421, 153

\bibitem[{{Stecker} {et~al.}(2007){Stecker}, {Baring}, \&
  {Summerlin}}]{stecker07}
{Stecker}, F.~W., {Baring}, M.~G., \& {Summerlin}, E.~J. 2007, \apjl, 667, L29

\bibitem[{{Tagliaferri} {et~al.}(2000){Tagliaferri}, {Ghisellini}, {Giommi},
  {Chiappetti}, {Maraschi}, {Celotti}, {Chiaberge}, {Fossati}, {Massaro},
  {Maesano}, {Montagni}, {Nesci}, {Nucciarelli}, {Pian}, {Raiteri},
  {Tavecchio}, {Tosti}, {Treves}, {Villata}, \& {Wolter}}]{tagliaferri00}
{Tagliaferri}, G., {Ghisellini}, G., {Giommi}, P., {et~al.} 2000, \aap, 354,
  431

\bibitem[{{Ulrich} {et~al.}(1997){Ulrich}, {Maraschi}, \& {Urry}}]{ulrich97}
{Ulrich}, M.-H., {Maraschi}, L., \& {Urry}, C.~M. 1997, \araa, 35, 445

\bibitem[{{Urry} \& {Padovani}(1995)}]{urry95}
{Urry}, C.~M. \& {Padovani}, P. 1995, \pasp, 107, 803

\bibitem[{{V{\'e}ron-Cetty} \& {V{\'e}ron}(2006)}]{veron-cetty06}
{V{\'e}ron-Cetty}, M. \& {V{\'e}ron}, P. 2006, \aap, 455, 773

\bibitem[{{Villata} {et~al.}(2000){Villata}, {Raiteri}, {Popescu}, {Sobrito},
  {De Francesco}, {Lanteri}, \& {Ostorero}}]{villata00}
{Villata}, M., {Raiteri}, C.~M., {Popescu}, M.~D., {et~al.} 2000, \aaps, 144,
  481

\bibitem[{{Virtanen} \& {Vainio}(2005)}]{virtanen05}
{Virtanen}, J.~J.~P. \& {Vainio}, R. 2005, \apj, 621, 313

\end{thebibliography}
\end{document}